\documentclass{article}
\usepackage{amsmath}

\begin{document}

\title{On states, channels and purification\thanks{Work partially supported by INTAS
grant 00-738.}}
\author{A. S. Holevo}
\date{}
\maketitle

\begin{abstract}
In this note we introduce purification for a pair $(\rho ,\Phi ),$ where $
\rho $ is a quantum state and $\Phi $ is a channel, which allows in
particular a natural extension of the properties of related information
quantities (mutual and coherent informations) to the channels with arbitrary
input and output spaces.
\end{abstract}

Given a state $\rho $ and a channel $\Phi $ there are three fundamental
entropy quantities: the input entropy $H(\rho )$, the output entropy $H(\Phi
\lbrack \rho ]),$ and the entropy exchange $H(\rho ,\Phi ),$ the last being
defined as $H((\Phi \otimes \mathrm{Id})[|\psi _{\rho }><\psi _{\rho }]),$
where $|\psi _{\rho }>$ is a purification of $\rho .$ From these entropies
one derives several information quantities of which the mutual information $
I(\rho ,\Phi )=H(\rho )+H(\Phi \lbrack \rho ])-H(\rho ,\Phi )$ and the
coherent information $I_{c}(\rho ,\Phi )=H(\Phi \lbrack \rho ])-H(\rho ,\Phi
)$ are of the main importance. The quantities $I,I_{c}$ have a number of
useful properties, the standard proof of which uses a dynamical
representation of the channel $\Phi $ via interaction of the system in
question with an environment (see e. g. \cite{AC}, \cite{NC}). Such a
representation presupposes that the system preserves its identity during the
interaction and hence the input and the output spaces of the channel are
necessarily identical too. However from the viewpoint of information theory
this is quite an unnatural restriction. One of the properties which we use
as an illustration is the data processing inequality for two channels $\Phi
_{1},\Phi _{2}:$
\begin{equation*}
I_{c}(\rho ,\Phi _{2}\Phi _{1})\leq I_{c}(\rho ,\Phi _{1}).
\end{equation*}
It is natural to expect that it holds for arbitrary channels satisfying the
only restriction that the output of the first channel is equal to the input
of the second. But using dynamical representations restrict us to the case
where both channels act in the same space. The present note removes this
restriction by introducing a new tool: a purification of the couple $(\rho
,\Phi ),$ from which all the three entropy quantities emerge with equal
status.

Consider the spectral decomposition
\begin{equation*}
\rho =\sum_{j=1}^{d}\lambda _{j}|e_{j}><e_{j}|,\quad \lambda _{j}\geq
0,\sum_{j=1}^{d}\lambda _{j}=1
\end{equation*}
in the input Hilbert space $\mathcal{H}_{in},$ and the Kraus decomposition
of the channel
\begin{equation*}
\Phi \lbrack \sigma ]=\sum_{\alpha =1}^{N}A_{\alpha }\sigma A_{\alpha
}^{\ast },\quad \sum_{\alpha =1}^{N}A_{\alpha }^{\ast }A_{\alpha }=I,
\end{equation*}
where $A_{\alpha }:\mathcal{H}_{in}\rightarrow \mathcal{H}_{out}.$ We
introduce the three systems $R,Q,E$ with the Hilbert spaces defined as
follows:
\begin{equation*}
\mathcal{H}_{R}=l_{d}^{2}\simeq \mathcal{H}_{in};\quad \mathcal{H}_{Q}=
\mathcal{H}_{out};\quad \mathcal{H}_{E}=l_{N}^{2},
\end{equation*}
where $l_{n}^{2}$ is the standard Hilbert space of $n$-dimensional vectors
and $\simeq $ denotes unitary equivalence. Consider the tensor product $
\mathcal{H}=\mathcal{H}_{R}\otimes \mathcal{H}_{Q}\otimes \mathcal{H}_{E}$
which can be realized as the space of vectors $|\psi >=\left[ \psi _{j\alpha
}\right] $ with the components $\psi _{j\alpha }\in \mathcal{H}_{out}.$ An
operator $X$ in $\mathcal{H}$ is represented by a square matrix $\left[
X_{k\beta }^{j\alpha }\right] $, the elements of which are operators in $
\mathcal{H}_{out}.$ Partial traces in $\mathcal{H}$ are computed according
to the formulas
\begin{equation*}
\mathrm{Tr}_{R}X=\left[ \sum_{j=1}^{d}X_{j\beta }^{j\alpha }\right] ;\quad
\mathrm{Tr}_{Q}X=\left[ \mathrm{Tr}X_{j\beta }^{j\alpha }\right] ;\quad
\mathrm{Tr}_{E}X=\left[ \sum_{\alpha =1}^{N}X_{k\alpha }^{j\alpha }\right] .
\end{equation*}
Take the unit vector
\begin{equation*}
|\psi _{(\rho ,\Phi )}>=\left[ \sqrt{\lambda _{j}}A_{\alpha }|e_{j}>\right]
\in \mathcal{H},
\end{equation*}
and the corresponding pure state $\Omega =|\psi _{(\rho ,\Phi )}><\psi
_{(\rho ,\Phi )}|,$ then the partial states are
\begin{equation*}
\Omega _{R}=\left[ \lambda _{j}\delta _{k}^{j}\right] \simeq \rho ;\quad
\Omega _{Q}=\Phi \lbrack \rho ];\quad \Omega _{E}=\left[ \mathrm{Tr}
A_{\alpha }\rho A_{\beta }^{\ast }\right] ,
\end{equation*}
therefore
\begin{equation*}
H(\Omega _{R})=H(\rho );\quad H(\Omega _{Q})=H(\Phi \lbrack \rho ]);\quad
H(\Omega _{E})=H(\rho ,\Phi ),
\end{equation*}
where the last equality follows from the facts that
$H(\Omega_{E})=H(\Omega_{RQ})$ and $$\Omega_{RQ}=[\sqrt{\lambda _{j}}
\sqrt{\lambda _{k}}\Phi[|e_j><e_k|]]
=({\rm Id}\otimes\Phi)[|\psi_{\rho}><\psi_{\rho}|],\;|\psi_{\rho}>
=[\sqrt{\lambda _{j}}|e_j>].$$ It also coinsides with
the well-known expression for the
entropy exchange in the dynamical case, see e. g. \cite{lin}.

Let us see how this can be used for the proof of the general data processing
inequality. Let $\left\{ A_{\alpha }\right\} ,\left\{ B_{\mu }\right\} $ be
the components of the Kraus decompositions for the channels $\Phi _{1},\Phi
_{2}.$ By using the purification $\Omega ^{1}=|\psi _{(\rho ,\Phi
_{1})}><\psi _{(\rho ,\Phi _{1})}|$ in $\mathcal{H}_{R}\otimes \mathcal{H}
_{Q_{1}}\otimes \mathcal{H}_{E_{1}}$ we obtain
\begin{equation*}
I_{c}(\rho ,\Phi _{1})=H(\Omega _{RE_{1}}^{1})-H(\Omega _{E_{1}}^{1}).
\end{equation*}
For the superposition $\Phi _{2}\Phi _{1}$ we use the purification
\begin{equation*}
|\psi _{(\rho ,\Phi _{2}\Phi _{1})}>=\left[ \sqrt{\lambda _{j}}B_{\mu
}A_{\alpha }|e_{j}>\right] \in \mathcal{H}_{R}\otimes \mathcal{H}
_{Q_{2}}\otimes \mathcal{H}_{E_{1}}\otimes \mathcal{H}_{E_{2}},
\end{equation*}
where $\mathcal{H}_{E_{2}}=l_{M}^{2},\quad M$ being the number of the
components in the Kraus decomposition of $\Phi _{2}.$ If $\Omega ^{12}$ is
the corresponding pure state, then
\begin{equation*}
I_{c}(\rho ,\Phi _{2}\Phi _{1})=H(\Omega _{RE_{1}E_{2}}^{12})-H(\Omega
_{E_{1}E_{2}}^{12}).
\end{equation*}
The data processing inequality will follow from strong subadditivity if we
show that $\Omega _{RE_{1}}^{1}=\Omega _{RE_{1}}^{12}.$ But
\begin{equation*}
\Omega _{RE_{1}}^{1}=\mathrm{Tr}_{Q_{1}}\Omega ^{1}=\left[ \sqrt{\lambda
_{j}\lambda _{k}}<\psi _{k}|A_{\beta }^{\ast }A_{\alpha }|\psi _{j}>\right] ,
\end{equation*}
while
\begin{equation*}
\Omega _{RE_{1}}^{12}=\mathrm{Tr}_{Q_{2}E_{2}}\Omega ^{12}=\left[ \sqrt{
\lambda _{j}\lambda _{k}}\sum_{\mu =1}^{M}<\psi _{k}|A_{\beta }^{\ast
}B_{\mu }^{\ast }B_{\mu }A_{\alpha }|\psi _{j}>\right] ,
\end{equation*}
which is indeed the same.

To see that the data processing inequality is not special, and
other properties can be treated in a similar way, let us establish
subadditivity of quantum mutual information
\begin{equation*}
I(\rho_{12},\Phi _{1}\otimes \Phi _{2})\leq I(\rho_{1},\Phi _{1})+I(\rho_{2},\Phi
_{2}).
\end{equation*}
Let $\Omega ^{12},\Omega ^{1},\Omega ^{2}$ be the purifying states for the
corresponding systems, constructed according to the recepee above. Then in
obviuos notations the inequality becomes
\begin{eqnarray*}
&&H(\Omega _{R_{12}}^{12})+H(\Omega _{Q_{1}Q_{2}}^{12})-H(\Omega
_{E_{1}E_{2}}^{12}) \\
&&\leq H(\Omega _{R_{1}}^{1})+H(\Omega _{Q_{1}}^{1})-H(\Omega
_{E_{1}}^{1}) \\&& + H(\Omega _{R_{2}}^{2})+H(\Omega
_{Q_{2}}^{2})-H(\Omega _{E_{2}}^{2}), \end{eqnarray*}
or, taking into
account purifications, \begin{eqnarray*} &&H(\Omega
_{Q_{1}Q_{2}E_{1}E_{2}}^{12})+H(\Omega _{Q_{1}Q_{2}}^{12})-H(\Omega
_{E_{1}E_{2}}^{12}) \\ &&\leq H(\Omega _{Q_{1}E_{1}}^{1})+H(\Omega
_{Q_{1}}^{1})-H(\Omega _{E_{1}}^{1}) \\&& + H(\Omega
_{Q_{2}E_{2}}^{2})+H(\Omega _{Q_{2}}^{2})-H(\Omega _{E_{2}}^{2}).
\end{eqnarray*}
This again will follow from repeated use of strong subadditivity and
from subadditivity of $H(\Omega _{Q_{1}Q_{2}}^{12})$ if we show that
\begin{equation*}
\Omega _{Q_{1}E_{1}}^{1}=\Omega _{Q_{1}E_{1}}^{12},\quad \Omega
_{Q_{2}E_{2}}^{2}=\Omega _{Q_{2}E_{2}}^{12}.
\end{equation*}
But the first equality just means that
\begin{equation*}
\left[ \sum_{\mu =1}^{M}\mathrm{Tr}_{Q_{2}}(A_{\alpha }\otimes B_{\mu
})\rho_{12}(A_{\beta }\otimes B_{\mu })^{\ast }\right] =\left[ A_{\alpha
}\rho_{1}A_{\beta }^{\ast }\right] ,
\end{equation*}
and similarly the second.

Let us make some general remarks. Unlike the case of dynamical
representation of the channels, where one considers the ``in'' systems $
R,Q,E $ and the ``out'' systems $R^{\prime },Q^{\prime },E^{\prime }$ etc.,
here we have only the ``out'' systems with $R$ being identical to the input
of $Q $ so there is no sense in ``primed'' notations. As in other cases, we
observe a kind of complementarity between statistical and dynamical aspects
of quantum description: our purification is ``less physical'' in that it
does not admit a dynamical interpretation, but it is more relevant to the
statistical structure related to the pair $(\rho ,\Phi )$ since it does not
involve any arbitrariness from outside.

\end{document}